\documentstyle[12pt,aps,prl]{revtex}

\begin{document}

\title
{Magnetization curves and ac susceptibilites
in type-II superconductors: geometry-independent similarity 
and effect of irreversibility mechanisms } 

\author
{G.M.Maksimova, D.Yu.Vodolazov, I.L.Maksimov}

\address
{Nizhny Novgorod University, Nizhny Novgorod 603600 Russia}

\maketitle

\begin{center}
keywords: edge barrier, bulk pinning, ac susceptibility
\end{center}

\begin{abstract}

The magnetic 
characteristics for superconducting samples of different
geometries (plates and films of various widths) and orientation with 
respect to external field are compared. A similarity is established 
between the magnetization curves $M(H)$ and the corresponding 
susceptibilities. A sign reversal is predicted for the 3rd 
harmonic of the ac susceptibillity when the irreversibility 
mechanism changes from bulk pinning to edge barrier. 
A mutual influence of the edge barrier and bulk pinning on the magnetization 
curves and field-dependent  ac susceptibility  is studied 
on the basis of an exactly solvable model (narrow plate in a parallel field 
and/or narrow film in a perpendicular magnetic field) .

\end{abstract}

\section{Introduction}

 Study of the magnetic characteristics of type-II superconductors and 
their behaviour in a magnetic field varying in time is of active 
interest from both the fundamental and applied standpoints. A large 
number of theoretical and experimental works \cite{1,2,3,4,5,6,7,8,9,10,11}
have been devoted to investigation into the magnetization of 
samples of various 
geometries, hysteresis losses and ac susceptibilities 
$\chi_n^{ac}=\chi_n'+i\chi_n''$. Two mechanisms are known to be 
responsible for these characteristics, specifically, bulk pinning 
and edge barrier that prevents vortex entry into (exit from) a 
smooth-surface sample. The effects produced by these mechanisms 
on the shape of the magnetization - and/or hysteresis curves have been 
considered by different authors \cite{2,4,8,9,10,11}. However, the influence 
of an edge barrier on field dependences of the hysteresis losses and 
the behaviour of higher harmonics of susceptibility have not been 
studied adequately so far. In particular,  as reported in our recent 
publication \cite{12}, the allowance for the edge barrier for a narrow 
film in a transverse magnetic field results in a sign  reversal of
the 3d harmonics of susceptibility $\chi_3'(H)$ and $\chi_3''(H)$. 
This is in stark contrast with  the situation in which bulk pinning 
is the major mechanism of irreversibility. Therefore, it would 
be extremely important to analyse a joint influence of an edge 
barrier and bulk pinning on the irreversible characteristics of type-II 
superconducting samples. For bulk superconductors a similar problem 
was studied in \cite{1} by Clem who has formulated a generalized local 
model of the critical state. This model provides an adequate basis 
for calculation of the magnetization curves and ac susceptibility  
for bulk superconductors \cite{13,14}. Yet, 
in low-dimensional superconductors (single crystals with high 
demagnetization factor, films) the effect 
of edge barrier and bulk pinning on magnetic characteristics is still 
not fully understood. For example, in \cite{15} the authors take account
of the joint influence of edge barrier and bulk pinning on the shape 
of the magnetization curve  for a wide film in a perpendicular magnetic
field. Unfortunately, the dependences M(H) obtained therein cannot 
be expressed in an analytical form, which does not allow to find the 
field and temperature dependences of magnetic susceptibility. It is, 
therefore, highly desirable to formulate an analytically solvable 
generalized model of the critical state for superconductors of 
arbitrary geometry, both longitudinal (as shown in Fig.1a) and 
transverse (see Fig.1b) with respect to the external magnetic field. 

In this work we report a study on the behavior of magnetization and 
the harmonics of ac susceptibility as a function of the amplitude 
of an applied low-frequency magnetic field. As an example, we used 
an exactly solvable model of a narrow film with an edge barrier. 
This model accounts for arbitrary bulk pinning which is characterized 
by a certain dependence of the depinning current density $j_p$ on magnetic 
field $H$. For  specific calculations, two model dependences $j_p(H)$ 
were used. One of them assumes that the depinning current density $j_p$
is a constant value independent of $H$ (the Bean model). 
In the other case the magnetic characteristics were obtained for 
the Kim-Anderson (KA) model dependence $j_p(H)$.

The paper is organized as follows. In Section II we formulate a 
generalized model of the critical state in a  narrow film/plate. 
On the basis of this model we obtain current densities and vortex
 distributions  in the mixed state taking into account both 
 edge barrier and bulk pinning as major  irreversibility
mechanisms in type-II superconductors. Two specific depinning
models were studied: the Bean model (see II.A) and the Kim-Anderson model
(see II.B). The  magnetization curves for a narrow film/plate are obtained 
and the harmonics of ac susceptibility are calculated in Sec. III.
In Sec. IV  a comparison is made of the magnetic 
characteristics for superconducting samples of different
geometries (plates and films of various widths) and orientation with 
respect to external field. A similarity is found out 
between the magnetization curves $M(H)$ and the corresponding 
susceptibilities. A sign reversal effect  is predicted for the 3d harmonic 
of ac susceptibillity, following a change of irreversibility mechanisms 
(from bulk pinning to edge barrier). In Sec. V we discuss the 
applicability conditions of the approach used. The conditions of the
barrier-to-pinning crossover in the temperature-dependent ac susceptibilities 
are analysed. A  brief summary of our results is presented in Sec. VI.

\section{Generalized model of the critical state}

 Let us consider a narrow plate (Fig.1a) or a film (Fig.1b) of width 
$W$ and thickness $d<\lambda$ ($W<\lambda$ for the plate, 
$W<\lambda^2/d$ for the film; $\lambda$ being the London penetration depth), 
which is infinite in the $x$-direction, placed in a magnetic field 
$H = (0,0,H)$. In view of the symmetry of the problem, the screening
current has one component $j_x=j(y)$, which satisfies the equation
(see Appendix)

$$
\frac{1}{C_0}\frac{dj}{dy}=H-n(y)\phi_0 . \eqno(1)
$$

Here $C_0=cW/(8\pi\lambda^2)$, $n$ is the vortex density, $\phi_0$ is 
the magnetic flux quantum,  
and the dimensionless coordinate $|y|\leq 1$ is measured in units $W/2$;
the origin of the coordinates is  assumed to be in the sample centre 
$(y = 0)$. Equation (1) can be derived   from the full version of the 
Maxwell-London equation (see e.g. ref.\cite{9}) by neglecting the nonlocal 
(integral) term.

 Since in the absence of transport current $n(y) = n(-y)$, and 
$j(y) = -j(-y)$, we may reduce our consideration to only one half 
of the film (for definiteness, the right-hand one, $y>0$). 
It is assumed that a superconductor is placed in a magnetic field 
which is first quasistatically increasing to some value $H_0$
and then decreasing down to $-H_0$. As is known, given an edge barrier 
the vortices penetrate a sample when the near-the-edge current density 
reaches some critical value $j_s$ (for  ideal surface $j_s$ coincides 
with the Ginzburg-Landau depairing current density $j_{GL}$).  
Therefore, up to the field $H=H_s = j_s/C_0$ the film remains in the 
Meissner state (see Fig. 2a):

$$
j(y)=yHC_0 , \quad   0<y<1 , \eqno(2)
$$  

 Further increase of the magnetic field will cause penetration of 
vortices that will move inside a superconductor until the Lorentz 
force $F = j\phi_0/c$ acting on each vortex becomes equal to the pinning 
force $F_p = j_p \phi_0/c$.

\subsection{Field-independent bulk pinning}

 First consider a model in which the depinning current density is 
independent of $H$ (the Bean model). The vortices that have penetrated 
into the superconductor at $H > H_s$ will be distributed with the density 
$n(y) = H/\phi_0$ in a region $a(H)<y<b(H)$, where $a(H) = H_p/H$ 
defines the depth of vortex penetration, $H_p = j_p/C_0$. Unlike 
the classical Bean model, the field of complete penetration ($a= 0$)
here is $H =\infty$, which  is a general property of the nonlocal model 
of the critical state both in bulk \cite{16,17} and quasi-two-dimensional
\cite{4,9} superconductors. The parameter 
$b(H)=1-(H_s-H_p)/H$ determines the size of the vortex-free region 
(of width $1 - b=(H_s-H)/H$), from which the vortices are "driven" 
by a strong near-edge current $j>j_p$. The corresponding 
distribution of the current density has the form (see Fig. 2b):

$$
j(y)=
\left \{ \begin{array}{lll}
 \displaystyle{yHC_0} & 0<y<a ,\\
 \displaystyle{j_p} & a<y<b ,\\
 \displaystyle{HC_0(y-1)+j_s} & b<y<1 .
\end{array} \right. \eqno(3)
$$

 We now start reducing the magnetic field from $H_0$ to $-H_0$. 
Until the current density in the vortex-occupied region 
$a(H_0)\leq y \leq b(H_0)$ has reached the critical value, 
$j=-j_p$, the distribution of vortices will be "frozen" and the 
distribution of current density will be described by the expression 

$$
j(y)=
\left \{ \begin{array}{lll}
 \displaystyle{yHC_0} & 0<y<a_0 ,\\
 \displaystyle{y(H-H_0)C_0+a_0H_0C_0} & a_0<y<b_0 ,\\
 \displaystyle{yHC_0+H_0C_0(a_0-b_0)} & b_0<y<1 ,
\end{array} \right. \eqno(4)
$$

where $a_0=a(H_0)$, $b_0=b(H_0)$. The corresponding distributions
$j(y)$, $n(y)$ are shown in Fig. 2c.

 At the flux-defreezing field $H_{df}=H_0(H_0-H_s-H_p)/(H_0-H_s+H_p)$, 
the current density $j(b_0)$ becomes equal to $-j_p$ and the vortices 
start moving towards the film edge. The magnetic flux related to 
the vortices will stay in the film until the vortex-exit field, 
$H_{ex}=H_0-H_s-H_p$, is reached. So, in the field interval 
$H_{ex}\leq H \leq H_{df}$ the current density and vortex 
distribution will have the following form (see Fig. 2d)

$$
j(y)=
\left \{ \begin{array}{llll}
 \displaystyle{yHC_0} & 0<y<a_0 ,\\
 \displaystyle{y(H-H_0)C_0+a_0H_0C_0} & a_0<y<b_1 ,\\
 \displaystyle{-j_p} & b_1<y<b_2 ,\\
 \displaystyle{yHC_0+H_0C_0(a_0-b_0)} & b_2<y<1 ; 
\end{array} \right. \eqno(5a)
$$
 
$$
n(y)=
\left \{ \begin{array}{llll}
 \displaystyle{0} & 0<y<a_0 ,\\
 \displaystyle{H_0/\Phi_0} & a_0<y<b_1 ,\\
 \displaystyle{H/\Phi_0} & b_1<y<b_2 ,\\
 \displaystyle{0} & b_2<y<1 ,
\end{array} \right. \eqno(5b)
$$

where $b_1=2H_p/(H_0-H)$, $b_2=(H_0-H_s-H_p)/H$. A further decrease in 
the magnetic field starting from $H_{ex}$ will cause the vortices exit 
from the film at $0 \leq H \leq H_{ex}$, as well (see Fig. 2e):

$$
j(y)=
\left \{ \begin{array}{lll}
 \displaystyle{yHC_0} & 0<y<a_0 ,\\
 \displaystyle{y(H-H_0)C_0+a_0H_0C_0} & a_0<y<b_1 ,\\
 \displaystyle{-j_p} & b_1<y<1 ;
\end{array} \right. \eqno(6a)
$$
 
$$
n(y)=
\left \{ \begin{array}{lll}
 \displaystyle{0} & 0<y<a ,\\
 \displaystyle{H_0/\Phi_0} & a<y<b_1 ,\\
 \displaystyle{H/\Phi_0} & b_1<y<1 .
\end{array} \right. \eqno(6b)
$$

 Note that at $H=0$ the remaning vortices occupy only the region 
$a_0 \leq y \leq b_1$ (see Fig. 2f). The edge current density in 
this case exceeds $-j_s$ (which prevents entry of antivortices into 
the film), therefore, with a further decrease of the field the 
distribution

$$
j(y)=
\left \{ \begin{array}{lll}
 \displaystyle{yHC_0} & 0<y<a_0 ,\\
 \displaystyle{y(H-H_0)C_0+a_0H_0C_0} & a_0<y<b_1 ,\\
 \displaystyle{yHC_0-H_0C_0(a_0-b_1)} & b_1<y<1 ;
\end{array} \right. \eqno(7a)
$$
 
$$
n(y)=
\left \{ \begin{array}{lll}
 \displaystyle{0} & 0<y<a_0 ,\\
 \displaystyle{H_0/\Phi_0} & a_0<y<b_1 ,\\
 \displaystyle{0} & b_1<y<1 ,
\end{array} \right. \eqno(7b)
$$

will be valid right down to the field
$H_s^{(-)}=(H_0-H_s-H_p)/2-\sqrt{(H_0-H_s-H_p)^2/4+H_0(H_s-H_p)}$, 
at which vortices of the opposite sign ('antivortices') will start 
entering into the film. In the field range $-H_0 \leq H \leq H_s^{(-)}$
(see Fig. 2g) one finds:

$$
j(y)=
\left \{ \begin{array}{llll}
 \displaystyle{yHC_0} & 0<y<a_0 ,\\
 \displaystyle{y(H-H_0)C_0+a_0H_0C_0} & a_0<y<b_1 ,\\
 \displaystyle{-j_p} & b_1<y<b_3 ,\\
 \displaystyle{yHC_0-(H+H_s)'_0} & b_3<y<1 ;
\end{array} \right. \eqno(8a)
$$
 
$$
n(y)=
\left \{ \begin{array}{llll}
 \displaystyle{0} & 0<y<a_0 ,\\
 \displaystyle{H_0/\Phi_0} & a_0<y<b_1 ,\\
 \displaystyle{H/\Phi_0} & b_1<y<b_3 ,\\
 \displaystyle{0} & b_3<y<1 .
\end{array} \right. \eqno(8b)
$$

where $b_3=1+(H_s-H_p)/H$.

\subsection{Field-dependent pinning (Kim-Anderson model)}

Let us find the distribution of vortices and a current in a film with 
a magnetic-field-dependent  bulk density of the depinning current 
$j_p(H)=C_0H_{k2}^2/(H_{k1}+|H|)$ (KA model), taking edge barrier 
into account. We shall follow the same procedure  as considered above. 
Thus, for the magnetic field increased to the magnitude $H_0$, $j(y)$ 
is determined by expressions (3), where 
$a(H)=H_{k2}^2/(H(H_{k1}+|H|))$, $b(H)=a(H)+(H-H_s)/H$. 
When the field is decreasing from $H_0$ to $H_{df}$, the current density 
and vortex distributions are similar to those in (4) with new values for
$a=a(H_0)$, $b=b(H_0)$, and the field $H_{df}$ being determined from 
the equation

$$
b_0(H-H_0)+a_0H_0+\frac{H_{k2}^2}{H_{k1}+|H|}=0
$$

Then, with the field decreasing from $H_{df}$ to $H_{ex}$, $j(y)$ 
and $n(y)$ are defined by the dependence (5) in which now we have

$$
b_1=\frac{H_{k1}}{H_0-H}(\frac{1}{H_{k1}+H_0}+\frac{1}{H_{k1}+|H|}) ,
\, b_2=\frac{H_0b_0+b_1(H-H_0)}{H},
$$

and field $H_{ex}$ is determined from the condition 
$b_2(H_{ex})=1$. If the field  amplitude $H_0$ satisfies the condition 
$H_0<\sqrt{2}H_{k1}$, then with a further decrease of the magnetic 
field down to $-H_0$ the evolution of $j(y)$, $n(y)$ distributions 
is described by formulae (6-8) in which $b_3$ now is expressed as

$$
b_3=\frac{H_s+H}{H}-\frac{H_{k2}^2}{H(H_{k1}+|H|)} ,
$$ 

and field $H_s^{(-)}$ satisfies  the equation

$$
H+H_0(a_0-b_1(H))+H_s=0 .
$$

 For the amplitudes that meet the condition $H_0>\sqrt{2}H_{k1}$, 
as can be easily shown, the dependence $b_1(H)$ posesses a minimum 
at some field $H_t^{(+)}=(\sqrt{2}-1)(H_0+\sqrt{2}H_{k1})>0$. 
In this case expressions (6) will hold only at $H \geq H_t^{(+)}$. 
A detailed examination shows that with a still further decrease of 
the field the distribution of the current density and vortices density 
in a film takes the form

$$
j(y)=
\left \{ \begin{array}{llll}
 \displaystyle{yHC_0} & 0<y<a_0 ,\\
 \displaystyle{y(H-H_0)C_0+a_0H_0C_0} & a_0<y<b^* ,\\
 \displaystyle{j_0(y)} & b^*<y<b_{2a} ,\\
 \displaystyle{-j_p(H)} & b_{2a}<y<1 ;
\end{array} \right. \eqno(9a)
$$
 
$$
n(y)=
\left \{ \begin{array}{llll}
 \displaystyle{0} & 0<y<a_0 ,\\
 \displaystyle{H_0/\Phi_0} & a_0<y<b_1 ,\\
 \displaystyle{(H_{k1}+H_{2a})(b^*-y)/2b^*\Phi_0+H_{2a}/\Phi_0} 
 & b^*<y<b_{2a} ,\\
 \displaystyle{H/\Phi_0} & b_{2a}<y<1 ,
\end{array} \right. \eqno(9b)
$$

where $b^*$ is determined from the condition $b^*=b_1(H_t^{(+)})$,
$b_{2a}$ from the equality $j_0(b_{2a})=-j_p(H)$, and $j_0(y)$ is 
defined as

$$
j_0(y)=C_0(H-H_t^{(+)})y-C_0\frac{H_{k1}+H_t^{(+)}}{2b^*}(b^*-y)^2+C_0b^*
(H_t^{(+)}-H_0)+a_0H_0C_0
$$       

 This current distribution is valid in the field range $0<H<H_t^{(+)}$.
If the field keeps decreasing, then, by analogy with the Bean model 
(provided $H_{s}>H_{k2}^2/H_{k1}$), antivortices at first will not be 
able to enter the film, and the current/vortex distribution in the film 
will be described by the following expressions

$$
j(y)=
\left \{ \begin{array}{llll}
 \displaystyle{yHC_0} & 0<y<a_0 ,\\
 \displaystyle{y(H-H_0)C_0+a_0H_0C_0} & a_0<y<b^* ,\\
 \displaystyle{j_0(y)} & b^*<y<b_{2a} ,\\
 \displaystyle{yHC_0-j_p(H)-b_{2a}HC_0} & b_{2a}<y<1 ;
\end{array} \right. \eqno(10a)
$$
 
$$
n(y)=
\left \{ \begin{array}{llll}
 \displaystyle{0} & 0<y<a_0 ,\\
 \displaystyle{H_0/\Phi_0} & a_0<y<b_1 ,\\
 \displaystyle{(H_{k1}+H_{2a})(b^*-y)/2b^*\Phi_0+H_{2a}/\Phi_0} 
 & b^*<y<b_{2a} ,\\
 \displaystyle{0} & b_{2a}<y<1 .
\end{array} \right. \eqno(10b)
$$

 With a further decrease of the magnetic field two situations are possible, 
depending on the problem parameters. In one case, at first $b_{2a}$ reduces 
to $b^*$ (which occurs at $H=H_t^{(-)}$), following which the current 
density at the edge reaches the antivortices-entry threshold $-j_s$. In the 
other case, the edge current density first falls down to $-j_s$ 
(it takes place at some field $H=\tilde H$), and then $b_{2a}$ 
decreases to $b^*$. For definiteness, we shall consider only the 
first situation, $|H_t^{(-)}|<|\tilde H|$. Here expressions (10) will 
hold right down to the field $H=H_t^{(-)}$ which is determined by 
the condition $b_{2a}=b^*$. Further, down to the field 
$H=-H_0$, $j(y)$ and n(y) are found from (7,8) with the $b_3$ and 
$H_s^{(-)}$ that have been introduced above. 

 Note that a similar method can be used to obtain the current/vortex 
density distribution for an arbitrary function $j_p(H)$. For a 
monotonically decreasing function $j_p(H)$ the resulting 
distributions will have a qualitative similarity to those obtained 
above. In the peak-effect conditions characterized by a nonmonotonic 
behaviour of $j_p(H)$ the issue on the form of $j(y)$ and $n(y)$ 
distributions needs to be addressed separately. 

\section{Magnetization curves and ac-susceptibility harmonics for 
narrow films}
	
\subsection{Magnetization curves}

 The above expressions for the current density allow to find the 
magnetization dependence for superconducting plates (films) in 
different ranges of field and obtain a hysteresis curve. For the 
Bean model given $H_0 \geq H_s+H_p$ the magnetization curve $M(H)$ 
is defined as follows

$$
M(H)=\gamma
\left \{ \begin{array}{lllll}
 \displaystyle{H_0(b_0^3-a_0^3)/2+3H_0(a_0-b_0)/2+H,} & H_{df}<H<H_0 ,\\
 \displaystyle{b_1^3(H_0-H)/2-H_0a_0^3/2+Hb_2^3/2+H+3H_0(a_0-b_0)/2,} 
 & H_{ex}<H<H_{df} ,\\
 \displaystyle{b_1^3(H_0-H)/2-H_0a_0^3/2+3b_1(H-H_0)/2+3H_0a_0/2,} 
 & 0<H<H_{ex} ,\\
 \displaystyle{H_0(b_1^3-a_0^3)/2+3H_0(a_0-b_1)/2+H,} & H_s^{(-)}<H<0 ,\\
 \displaystyle{b_1^3(H_0-H)/2-H_0a_0^3/2+H-3(H+H_s)/2+Hb_3^3/2,} 
 & -H_0<H<H_s^{(-)} , 
\end{array} \right. \eqno(11)
$$

where $\gamma=-C_0W/12c$.

 At a relatively low field magnitude, $H_0<H_s+H_p$, the mode described 
by expressions (4) will immediately change for the mode (7). 
Thus, the dependence $M(H)$ in the fields interval $H_s<H_0<H_s+H_p$ 
will have the form

$$
M(H)=\gamma
\left \{ \begin{array}{llll}
 \displaystyle{H_0(b_0^3-a_0^3)/2+3H_0(a_0-b_0)/2+H,} & H_{df}<H<H_0 ,\\
 \displaystyle{H_0(b_1^3-a_0^3)/2+3H_0(a_0-b_1)/2+H,} 
 & H_s^{(-)}<H<H_{df} ,\\
 \displaystyle{b_1^3(H_0-H)/2-H_0a_0^3/2+H-3(H+H_s)/2+Hb_3^3/2,} 
 & -H_0<H<H_s^{(-)} ,
\end{array} \right. \eqno(12)
$$

In Fig. 3 the dependence $M(H)$ is shown for different values of fields 
$H_p$ and $H_s$. It is seen that the edge barrier causes peaks to arise 
on the magnetization curve $M(H)$ at $H \approx \pm H_s$. With an 
increase of the ratio $H_s/H_p$ the amplitude of the peak grows and the 
latter shifts towards higher fields. Such a behavior of magnetization 
is similar to the corresponding dependence reported for wide films in 
\cite{15}.

Within the Kim-Anderson model, if $H_s>H_{k2}^2/H_{k1}$,
$H_0>H_s+H_{k2}^2/H_{k1}$, $H_0>\sqrt{2}H_{k1}$, $|H_t^{(-)}|<|\tilde H|$,
the dependence $M(H)$ is written in the form

$$
M(H)=\gamma
\left \{ \begin{array}{lllllll}
 \displaystyle{H_0(b_0^3-a_0^3)/2+3H_0(a_0-b_0)/2+H,} & H_{df}<H<H_0 ,\\
 \displaystyle{b_1^3(H_0-H)/2-H_0a_0^3/2+Hb_2^3/2+H+3H_0(a_0-b_0)/2,} 
 & H_{ex}<H<H_{df} ,\\
 \displaystyle{b_1^3(H_0-H)/2-H_0a_0^3/2+3b_1(H-H_0)/2+3H_0a_0/2,} 
 & H_t^{(+)}<H<H_{ex} ,\\
 \displaystyle{Ha_0^3+(H-H_0)((b^*)^3-a_0^3)+3H_0a_0((b^*)^2-a_0^2)/2-} & \\ 
 \displaystyle{3H_{k2}^2(1-b_{2a}^2)/2(H_{k1}+|H|)+m_0,} & 0<H<H_t^{(+)} ,\\
 \displaystyle{Ha_0^3+(H-H_0)((b^*)^3-a_0^3)+3H_0a_0((b^*)^2-a_0^2)/2-} & \\
 \displaystyle{3H_{k2}^2(1-b_{2a}^2)/2(H_{k1}+|H|)+m_0+Hb_{2a}^3/2+
 H(1-3b_{2a}/2),} & {H_t^{(-)}}<H<0 ,\\
 \displaystyle{H_0(b_1^3-a_0^3)/2+3H_0(a_0-b_1)/2+H,} 
 & H_s^{(-)}<H<H_t^{(-)} ,\\
 \displaystyle{b_1^3(H_0-H)/2-H_0a_0^3/2+H-3(H+H_s)/2+Hb_3^3/2,} 
 & -H_0<H<H_s^{(-)} ,
\end{array} \right. \eqno(13)
$$

where $m_0=6/('_0W)\int_{b^*}^{b_{2a}}yj_0(y)dy$.

When $H_0<\sqrt{2}H_{k1}$, the dependence $M(H)$ will be fully 
identical to (11) with the appropriate values of parameters 
$a,b_1,b_2,b_3,H_{df},H_{ex},H_s^{(-)}$ obtained using the KA
approach. 

 Fig.4 shows dependence (13) for different values of the parameters 
related to  the Kim-Anderson depinning model and edge barrier. Unlike 
in the Bean model, the magnetization curve here has a maximum even 
without an edge barrier. Given an edge barrier, the $M(H)$ peak 
amplitude increases and the peak itself shifts towards the higher fields.

\subsection{ac - susceptibility}

 If an external magnetic field H changes harmonically in time, 
$H(t)=H_0\cos(\omega t)$, magnetization $M(t)$ is also a periodical 
function with a period $T=2\pi/\omega$, whose Fourier components of 
the magnetization $M(H)$ determine the ac-susceptibility harmonics

$$
\chi_n^{ac}=-\frac{2\gamma}{\pi H_0}\int \limits _{0}^{\pi}M(\Theta)
e^{in\Theta}d\Theta=\chi_n'+i\chi_n''. \eqno (14)
$$

where $\Theta=\omega t$.

 It follows from the symmetry of the problem ($M(H)=-M(-H)$) that 
all of the even harmonics are zero. Fig. 5a-5b shows the imaginary 
and the real parts of the first (i.e. fundamental) and third 
harmonics of ac susceptibility, 
calculated by formulae (13) both with- and without an edge barrier 
(curves 1 and 2, respectively). One can see that existence of an 
edge barrier produces a quantitative (first harmonic) and a 
qualitative (change of sign in the real part of the 3rd harmonic) 
effect on the form of the dependence $\chi_n(H_0)$. We believe that 
the sign reversal effect (SRE) which was for the first time predicted by 
the authors in \cite{12} is a vivid manifestation of how an edge 
barrier affects the $\chi_3(H_0)$ behavior. It should be mentioned 
that a similar behavior is typical for a specific set of higher 
harmonics, as well:  $\chi_{4k-1}'(H_0)$, $\chi_{4k-1}''(H_0)$ ($k=2,3$).
Note also that the SRE is not observed for the harmonics
$\chi_{4k+1}'(H_0)$, $\chi_{4k+1}''(H_0)$ ($k=1,2$).

\section{Geometry effect on magnetic characteristics of
type-II superconductors}

 As follows from our analysis, the magnetic characteristics of
various-width plates ($W \leq \lambda$ and $W \gg \lambda$ ) in a
parallel magnetic field and also of both narrow ($W \leq \lambda^2/d$)
and wide ($W \gg \lambda^2/d$) films in a transverse magnetic field
are quite similar. The analogous similarity properties of the
magnetization curves were pointed out by \cite{18}  where  curves
$M(H)$ were compared for macroscopic samples of
different geometries posessing no edge/surface 2barrier .

 In Fig. 6a the magnetization curves are shown for samples without a
bulk pinning. Curve 1 was obtained for a narrow film (formula (11)
with $H_p=0$), curve 2 corresponds to a wide plate in a parallel field
(as calculated within the nonlocal model \cite{16}),
 curve 3 is for a wide film in a transverse magnetic field \cite{8},.
The magnetic field is measured in units of $H_s$, essentially the
first-vortex entry field which is different in either case:\\
1)$H_s=\Phi_0/2\pi\xi W$ \cite{19,20},
2)$H_s=\Phi_0/2\pi\xi\sqrt{W\lambda^2/d}$ \cite{19},\cite{21},
3)$H_s=\Phi_0/2\sqrt{2}\pi\lambda\xi$ \cite{22}.

 The magnetization is normalized so as to ensure the condition
$M(H)=H$ in the Meissner state. As is seen from Fig.6a,
the dependences $M(H)$ in the appropriate units are qualitatively alike,
therefore, the magnetization harmonics $\chi_n^{ac}(H_0)$ (Fig. 7a-10a)
also feature a similarity. (Note that this property is exhibited
by susceptibilities with $n>3$, as well).

 A second set of figures, 6b-10b, describes the situations in which 
account is taken only of a bulk pinning ($j_p(H) = const$), and 
an edge barrier is neglected. Curves 1 correspond to narrow films 
(see dependence (11) at $H_s=H_p$), curves 2 are for wide 
films in the longitudinal geometry \cite{16}, curves 3 were obtained
for wide films in the transverse geometry \cite{4} . The magnetic field is 
measured in units of $H^*$ which for narrow films is equal to $j_p/C_0$,
for curves 2 - $H^*=2\pi j_pW/c$, for curves  3 - $H^*=4j_pd/c$ . 
Magnetization is measured in units of $j_pW/8c$. Note that in all cases 
considered the curves $M(H)$ never saturate, which is a direct consequence 
of the nonlocality effect. A similarity 
between the corresponding dependences $M(H)$ obtained in different 
geometries of the problem is more conspicuous here than in the case 
of a zero-bulk-pinning (see above).  Besides, it is obvious that the 
third harmonic (both the imaginary and the real parts) changes sign 
in transition from the edge-barrier to bulk-pinning mode. 

Performing simple calculations
(for a solvable case of narrow films/plates) 
under condition $j_p \ll j_s$  one obtains
the following expressions for $\chi_3'(H_0)$ and $\chi_3''(H_0)$
in the limit $H_0 \gg H_s$ 

$$
\chi_3'(H_0 \gg H_s)=-\frac{2\gamma}{\pi H_0} \left( 4H_p\sqrt{
\frac{2H_p}{H_0}}-\frac{3\pi H_s^2}{4H_0} \right) , \eqno(15a)
$$

$$
\chi_3''(H_0 \gg H_s)=-\frac{2\gamma}{\pi H_0} \left(
\frac{3H_s^2}{2H_0}\ln\frac{H_0}{H_s}-H_p \right) . \eqno(15b)
$$

As follows from above expressions in the case of a superconductor without
bulk pinning  $\chi_3'(H_0\gg H_s)>0$ and $\chi_3''(H_0\gg H_s)<0$.
However, presence of bulk pinning leads to a sign reversal of
$\chi_3'(H_0)$ for sufficiently  large amplitudes
$H_0 > H_s(j_s/j_p)^3$  and of $\chi_3''(H_0)$ for amplitudes
$H_0 > H_s(j_s/j_p)\ln(j_s/j_p)$.

We wish to point out that for large amplitudes of the 
magnetic field $H_0 $, the numerical  results related to the  case of a
narrow film/plate coincide with the analytically obtained  asymptotics
(15 a,b) for $\chi_3'(H_0)$ and $\chi_3''(H_0)$. 
For an  additional control of the calculation accuracy
  we have also employed  a twice smaller  numerical step; 
the results practically have not changed
 (the difference did not exceed  $3 \%$ ).
In our opinion, the above arguments fully
confirm reliability of our  calculations.

 The third situation we have examined corresponds to the $j_p(H)$ by
the Kim-Anderson model (ignoring edge barrier). Magnetization curves
and ac-susceptibilities are shown in Figs 6c-10c. Here we consider
only one case for wide films, specifically, the longitudinal geometry,
since there are no analytical expressions to describe the dependence
$M(H)$ for wide films in the transverse geometry. In \cite{23} the 
results of numerical calculations of the M(H) dependence are reported
for a wide film. A similar dependence is obtained (given the appropriate 
choice of parameters) also for a wide plate in a parallel field. 
Curves 1 in Fig. 6c-10c correspond to a narrow plate (film), 
curves 2 are for a wide plate. The magnetic field is measured in 
units of $H^*$ which is equal to $j_p(0)/C_0$ for narrow films and to 
$2\pi j_p(0)W/c$ for wide films. In both the cases magnetization is 
measured in units of $j_p(0)W/8c$. The dependence $M(H)$ corresponding
to curve 1 (Fig.6c) is obtained by substitution of the expression 
$H_s=H_{k2}^2/(H_{k1}+|H|)$ in (13), while $M(H)$ for curve 2 shown 
in the same figure is obtained using the critical state model for a 
wide plate, in the same way as it was done in \cite{5,13} for an 
infinite cylinder in a parallel magnetic field. 
Note a practically absolute similarity of the curves at $H_0 \gg H^*$;
however, at lower $H_0$ the differences become noticeable. Therefore, 
at large amplitudes $H_0$ the susceptibility harmonics for these two cases 
almost coincide, whereas at small amplitudes they differ in quantity. 

We should emphasize that  the KA model also demonstrates
sign reversal in the third harmonic of ac susceptibility
(only in the real part, though) in transition from the edge-barrier to
bulk-pinning irreversibility mode. As is shown
by numerical analysis of narrow samples, in a physically
interesting case $H_0>H_{k1}$ this transition occurs when
$H_{k2}^2/H_{k1}>2H_{s}$. The latter
corresponds to the condition that the depinning current density
in a zero magnetic field $j_p(0)$ be twice higher than the density
of the edge-barrier suppression current. Note that such a situation
is practically impossible in superconductors with an ideal barrier
($j_s=j_{GL} \approx 10^8 A/cm^2$). Therefore, the edge barrier effect
has to be taken into account in study of the magnetic characteristics
of type II superconductors and of higher harmonics of ac susceptibility.

\section{Discussion}

 Let us discuss the applicability conditions for a generalized model
of critical state. A continuum approximation to describe vortex
distribution in narrow samples is valid, if $W \gg \xi$ (here $\xi$ is
the coherence length). Introduction of the function $n(y)$ implies
essentially the averaging of vortex distribution on a scale exceeding
the intervortex distance $\sim n^{-1/2}$. It is easily shown that this
averaging procedure is correct in the field range $\Phi_0/W^2
< H \ll \Phi_0/\xi^2$. For example, the applicability condition of
this model ($\xi \ll W < \lambda^2/d$) was met near the critical
temperature ($T \to T_c$) for a $Sn$ film in the experiment \cite{24}. 
Similar conditions can be provided also in $HTSC$ films based, 
for instance, on $YBaCuO$ with parameters $\lambda \sim 0.2 \mu m$ 
and $d \sim 0.1 \mu m$ at $T \approx 77 K$. 

 The results for magnetic susceptibilities, obtained in this work are
valid on condition of quasistatical variation of an external field.
The slowness of the magnetic field variation implies that there is
sufficient time for  a quasiequilibrium distribution of current and
vortex density to set in a sample, and the viscous losses related to
vortex motion can be neglected. As follows from formula (14), in this
case $\chi_n^{ac}$ is frequency independent. Absence of frequency
effects on magnetic susceptibilities was reported in refs \cite{9,14}
in the frequency domain $\omega/2\pi=30 \div 1000 Hz$ and at
amplitudes $H_0= 0.1 \mu T \div 1 mT$.

 Consider now the temperature dependence of $\chi_n^{ac}$. Note that 
the similarity described in Section 4 is mostly featured 
by field dependences $\chi_n^{ac}(H_0)$. The temperature dependences 
of $\chi_n(T)$ will differ appreciably for samples of different 
geometries and orientation.
This is predetermined by a different character of the 
temperature dependence of fields, $H_s(T)$ and $H_p(T)$ in wide and
narrow samples. Nevertheless, it is quite possible to observe a 
 sign reversal on the temperature dependence of $\chi_3'(T)$ due 
to  edge barrier effect. Indeed, let $H_p(T)\sim (1 - T/T_c)^\alpha$, 
and $H_s \sim (1 - T/T_c)^\beta$ and, besides, $\alpha>\beta$ (see e.g. 
\cite{3,24}). Then, even if at low temperatures $H_p(T) \gg H_s(T)$ 
(bulk pinning dominates), at $T\sim T_c$ 
$H_p(T) \ll H_s(T)$, i.e., the edge barrier mechanism of 
irreversibility will dominate. Such an interchange of the irreversibility
mechanisms results in a sign reversal of $\chi_3'$, which takes place 
at a certain crossover temperature $T^*$. 
Note that a similar crossover of magnetization was described in \cite{25}.

 It should be noted that a change of irreversibility 
mechanism may be followed by a  sign reversal not only in $\chi_3'(H_0)$,
as is demonstrated by the KA model (see Fig. 9a,9c), but also 
in $\chi_3''(H_0)$ (within the Bean model; see Fig. 9a, 10a-9b, 10b). 
This brings up a question about how sensitive the $\chi_3(H_0)$ behavior 
is to a bulk-pinning mechanism. We have considered the dependence 
$j_p(H)$ which corresponds to a collective mechanism of vortex 
pinning \cite{26} in the form

$$
j_p(H)=
\left \{ \begin{array}{ll}
 \displaystyle{j_{p0}\sqrt{H_k/H}} & |H|>H_k ,\\
 \displaystyle{j_{p0}} & |H|<H_k .
\end{array} \right. \eqno(16)
$$

 The behavior of $M(H)$ and $\chi_n^{ac}(H_0)$, found using Eq.(16) 
agrees qualitatively with the curves shown in Figs. 6c-10c (the KA model). 
Apparently, these data should also be valid qualitatively for an arbitrary 
function $j_p(H)$ monotonically decreasing with H. 
Therefore, we may conclude that the sign reversal effect in the real
part of the third susceptibility harmonics, following a change of 
irreversibility mechanism, is a general property of the mixed state, 
which is only slightly affected by specific features of edge barrier

Let us discuss now an unexpected  similarity of the magnetization curves
and ac susceptibilities. Indeed, in order to  calculate the
magnetic characteristics of a specific sample one should solve a 
second-order differential equation (see (A7) in the Appendix) 
(for a bulk superconductor case), the integral Bio-Savart equation (A5) 
(for a wide film in a  perpendicular geometry) and/or the first-order 
differential equation (A6) or (A8) (for a narrow film/plate). Surprizingly, 
the magnetization curves obtained on the basis of the solutions of these 
apparently different equations look quite similar (on condition that the 
same  irreversibility mechanism, i.e. an edge barrier or a specific type of 
the fluxoid pinning by the lattice defects, is employed) \cite{27} . 
Such geometry-independent similarity  obviously follows from the fact that 
current densities and vortex distributions in a narrow superconducting 
film/plate (see Fig. 2) and in  samples of other geometries are 
qualitatively similar in corresponding ranges of the magnetic field.
The difference in samples geometry and size results in  mere quantitative
modifications of the characteristic parameters (e. g. $a, b$, $H_{df}, H_{ex}$
etc.; for specific examples  see Refs. \cite{15,16,17} )
and in specific dependences $j_x(y)$ ¨ $n(y)$ corresponding 
to a sample of selected geometry. Thus, taking into account the 
above results, a narrow film/plate may be suggested as a basic 
model to study qualitative features of the magnetic properties of 
practically used type-II superconductors.

The geometry-related similarity revealed here seems to be caused by 
the one-parametric description of each irreversibility mechanism by 
means of introduction of a phenomenological current density for the 
barrier suppression $j_s$ or for the depinning $j_p(H)$. The microscopic 
basis of such universality still awaits a more detailed study.

\section{Summary}

 In this work the magnetic characteristics of superconducting
plates and films of various widths in an external quasistatistical
magnetic field $H=H_0\cos(\omega t)$ have been calculated. The study
was carried out within a generalized model of the critical state, which
accounts for both edge barrier and bulk pinning. To demonstrate the
effect of a surface (edge) barrier, particular models were considered,
describing individual influence of either irreversibility
mechanism (edge barrier or bulk pinning) on the magnetization and
ac-susceptibility of type II superconductors. The obtained results
have led us to the following conclusions: \\
1. The magnetic characteristics of type II superconductors are determined
by the type of irreversibility mechanism (bulk pinning or surface/edge
barrier) rather than geometrical parameters of samples (plates, films)
and their orientation relative to an external magnetic field. The latter
factors bring about only quantitative changes in the dependences
$M(H)$, $\chi_n^{ac}(H_0)$, which thus prove to be similar (for a
specified mechanism of vortex depinning).
2. A sign change effect in the real part of the third (7-th, 11-th)
harmonics of the magnetic susceptibility is predicted to follow a
change of irreversibility mechanism. A generalized model of the
critical state for narrow films has been used to determine the
conditions at which a crossover of the  $\chi_3^{ac}(H_0)$ dependence
takes place.

\section{Acknowledgments}

Authors wish to thank Prof. J. Clem and Prof. L.M. Fisher
for helpful discussions of the results obtained.
This work is supported by the Science Ministry of Russia (Project
No. 98-012). Partial support of the International Center for
Advanced Studies (INCAS) through Grant No.99-2-03 is gratefully
acknowledged.

\section{Appendix}

Let us consider  thin superconducting film in a perpendicular magnetic 
field (see fig. 1b) in a mixed state. The relation 
between  current density and vector potential in the London model reads

$$
{\bf j}=-\frac{c}{4\pi \lambda^2}({\bf A}-\frac {\phi_0}{2\pi} \nabla 
\varphi). \eqno (A1)
$$

Here $\varphi$ is a phase of the order parameter satisfying the equation

$$
{\bf \nabla} \times {\bf \nabla}\varphi ({\bf r})=
2\pi \delta ({\bf r} - {\bf r} '), \eqno (A2)
$$

where ${\bf r}'=(x', y')$ is the vortex coordinates.  The Ampere law 
 (in a gauge  ${\bf \nabla \cdot  A}=0$) reads

$$
\Delta {\bf A} = -\frac {4\pi} {c} {\bf j}. \eqno (A3)
$$

By employing a Green function for the three-dimensional Laplace operator 
$ \Delta $ we rewrite (A3) in the integral form

$$
{\bf A}({\bf r})={\bf A}_0({\bf r})+\frac{1}{c}\int \int \int \frac{{\bf j}
({\bf r '})} {\bf |r-r'|} dx'dy'dz', \eqno (A4)
$$

where ${\bf A}_0({\bf r})$ is the vector potential of the 
applied magnetic field ${\bf H}= {\bf \nabla \times  A}_0$;
the integration  in (A4) is performed over entire sample.  
With the help of Eqs. (A1)-(A4) we derive the Maxwell-London equation for 
the average current density $j(y)= d^{-1} \int^{d/2}_{-d/2} j_x(y,z) dz$ 
in a thin-film limit $d \ll \lambda $

$$
\frac {8\pi\lambda _{eff}}{cW} \frac {dj(y)}{dy}+\frac{2}{c}
\int_{-1}^{1} \frac {j(y')} {y'-y} dy' = d^{-1} [ H-n(y)\phi_0]. \eqno (A5)
$$

Here $n(y)$ is the vortex density  and distance is scaled in units 
of $W/2 $. Equation (A5) was derived for the first time in Ref. 
\cite {28}. For the case of a narrow film $\lambda_{eff}/W \gg 1$ 
when the  self-field of currents is neglected Eq. (A5) reduces to 

$$
\frac{4\pi\lambda^2}{c}\frac{dj_x}{dy} = H - n(y)\phi_0. \eqno (A6)
$$

The numerical solution of the equation (A5) shows that the integral term
becomes insignificant for sufficiently narrow films $W \leq \lambda_{eff}$. 
In the opposite case of wide films, the differential term in (A5) can be 
neglected (everywhere inside the  film except for areas near edges).
In the latter limit Eq. (A5) reduces to the canonical version of 
the Bio-Savart equation that is conventionally studied in a 
quasi-two-dimensional situation \cite{29,30}.

Consider now the mixed state of a superconducting plate in a 
parallel magnetic field (see fig. 1 ). By taking ${\rm curl} $ from 
both parts of the London relation (A1) with account for the
relation (A2) and Maxwell equations ${\bf \nabla \times H}=4\pi{\bf j}/c$ 
and ${\bf \nabla \cdot B}=0$ we obtain the nonlocal  equation 
for the distribution of a magnetic field (see \cite {16,17})

$$
H_l - \lambda^2 \frac{d^2 H_l}{dy^2} = n(y) \phi_0, \eqno (A7)
$$

where $H_l$ is a $z$-component of a local magnetic field 
${\bf H}_l=(0,0,H_l)$ (all other components vanish due to the 
symmetry of the problem). 
In the limiting case $\lambda \to 0$ (A7) reduces to the conventional
relation $H_l=n \phi_0$.

In case of narrow plates $ \lambda/W \gg 1 $ the differential term in 
(A7) is dominant and, moreover, $H_l \sim H$ since
the self-field of currents can be neglected. Employing relation 
$dH_l/dy=4\pi j_x/c $, valid for longitudinal geometry, we obtain 
the equation for the current density $j_x $

$$
 \frac{4\pi\lambda^2}{c}\frac{dj_x}{dy}=H-n(y)\phi_0. \eqno (A8)
$$

Thus, the equations describing the current density distribution 
for narrow films (see (A6)) and  for narrow plates (see (A8)) 
are equivalent. The mathematical equivalence of these equations follows 
from the fact that the current-induced self field may be neglected in these 
two extreme cases.

\newpage

{\bf Figure captions}

Fig.1

Problem geometry: a) plate in a parallel magnetic field;
b) film in a perpendicular magnetic field.

Fig.2

Current densities and vortex distributions in a 
narrow superconducting film/plate at different magnetic fields $H$.
a) $0<H\uparrow<H_s$, b) $H_s<H\uparrow<H_0$,
c) $H_{df}<H\downarrow<H_0$, d) $H_{ex}<H\downarrow<H_{df}$,
e) $0<H\downarrow<H_{ex}$, f) $H_s^{(-)}<H\downarrow<0$,
g) $-H_0<H\downarrow<H_s^{(-)}$ \\ 
$\uparrow$ indicates increasing field
and $\downarrow$ indicates decreasing field.

Fig.3

Magnetization curves for a narrow thin film with an edge barrier
and bulk pinning (Bean model): 1) $j_p=j_s$, 2) $j_p=0.6j_s$,
3) $j_p=0.3j_s$, 4) $j_p=0$.

Fig.4

Magnetization curves for a narrow thin  film with an
edge barrier and bulk pinning (Kim-Anderson model): 1) $j_p(0)=j_s$,
2) $j_p(0)=0.6j_s$, 3) $j_p(0)=0.3j_s$.

Fig.5

Dependence of magnetic susceptibilities on the amplitude of a quasistatical
applied magnetic field ($H=H_0\cos(\omega t)$) of a narrow-film 
superconductor (Kim-Anderson model of bulk pinning:  
$j_p(0)/j_s=1/5$, $H_{k1}=H_s$)).\\
Curve  1 - a sample with only bulk pinning. \\
Curve  2 - a sample with edge barrier and bulk pinning.\\
a) $\chi_1'(H_0)$, b) $\chi_1''(H_0)$, c) $\chi_3'(H_0)$,
d) $\chi_3''(H_0)$.

Fig.6

Magnetization curves for superconductors of various widths and geometries
with single mechanism of irreversibility.\\
a) - superconductor with edge barrier, b) superconductor with 
bulk pinning (Bean model),
c) - superconductor with bulk pinning (Kim-Anderson model). \\
Curve 1 - narrow film (plate), curve 2 - wide plate, curve 3 - wide film.

Fig.7

Dependences of the real part of the first harmonic of magnetic
susceptibility on the amplitude of an applied magnetic field $H_0$. \\
a) - superconductor with edge barrier, b) superconductor with bulk 
pinning (Bean model),
c) - superconductor with bulk pinning (Kim-Anderson model). \\
Curve 1 - narrow film (plate), curve 2 - wide plate, curve 3 - wide film.

Fig.8

Dependences of the imaginary part of the first harmonic of magnetic
susceptibility on the amplitude of an applied magnetic field $H_0$. \\
a) - superconductor with edge barrier, b) superconductor with bulk 
pinning (Bean model),
c) - superconductor with bulk pinning (Kim-Anderson model). \\
Curve 1 - narrow film (plate), curve 2 - wide plate, curve 3 - wide film.

Fig.9

Dependences of the real part of the third harmonic of magnetic
susceptibility on the amplitude of an applied magnetic field $H_0$. \\
a) - superconductor with edge barrier, b) superconductor with bulk 
pinning (Bean model),
c) - superconductor with bulk pinning (Kim-Anderson model).  \\
Curve 1 - narrow film (plate), curve 2 - wide plate, curve 3 - wide film.

Fig.10

Dependences of the imaginary part of the third harmonic of magnetic
susceptibility on the amplitude of an applied magnetic field $H_0$. \\
a) - superconductor with edge barrier, b) superconductor with bulk 
pinning (Bean model),
c) - superconductor with bulk pinning (Kim-Anderson model). \\
Curve 1 - narrow film (plate), curve 2 - wide plate, curve 3 - wide film.


\begin{thebibliography}{30}
\bibitem{27} In particular case, corresponding to
the absence of edge barrier, the analogous similarity of magnetic properties 
has been demonstrated by Gilchrist \cite{18} within the Bean model of 
the critical state.
\end{thebibliography}
\end{document}